\documentclass[conference]{IEEEtran}
\IEEEoverridecommandlockouts

\usepackage{amsmath,amssymb}
\usepackage{booktabs}
\usepackage{graphicx}
\usepackage{cite}
\usepackage{xspace}
\usepackage{microtype}
\usepackage{array}
\usepackage{tabularx}
\usepackage{xcolor}
\usepackage{tikz}
\usepackage[table]{xcolor}
\usepackage{balance}
\usetikzlibrary{arrows.meta,positioning}
\definecolor{lightblue}{RGB}{225,240,255}

\newcommand{\method}{ConQUESTA\xspace}
\newcommand{\RT}{R_{\mathrm{T1}}}
\newcommand{\Aval}{A_{\mathrm{val}}}
\newcommand{\pp}{\ensuremath{\,\mathrm{pp}}}
\newcolumntype{Y}{>{\centering\arraybackslash}X}

\begin{document}

\title{ConQUESTA: Topology-Aware Post-Training Structural Pruning of Quantum Neural Networks}

\author{\IEEEauthorblockN{Saim Rehman, Muhammad Shafique}
\IEEEauthorblockA{eBRAIN Lab, Division of Engineering, New York University Abu Dhabi (NYUAD)\\
Abu Dhabi, United Arab Emirates\\
sr7849@nyu.edu, muhammad.shafique@nyu.edu}}

\maketitle

\begin{abstract}
Deploying quantum neural networks (QNNs) on sparse hardware requires preserving task accuracy while reducing compiler-induced two-qubit gates, SWAPs, and depth. We present ConQUESTA (\underline{C}onstrained \underline{Qu}antum \underline{E}ntangler \underline{S}election with \underline{T}opology \underline{A}wareness), a post-training pruning framework that combines task saliency with compiler cost, screens candidate entanglers, and searches complete pruning masks under an explicit accuracy constraint. Unlike one-shot ranking, each mask is transpiled as a whole across multiple coupling topologies. Across 4-, 6-, and 8-qubit QNNs and eight model seeds at 55\% pruning, ConQUESTA achieves mean resource reductions of 61.6\% on Breast Cancer, 72.0\% on Digits01, and 82.2\% on Iris, exceeding the strongest matched simple baseline by 3.5, 1.7, and 8.9 percentage points, respectively. Fixed-mask recovery raises feasibility from 16/24 to 22/24 runs. In a fixed-k=4 scaling study on a scaling probe of 4 to 12-qubits, the raw mask space grew from 495 to 12.1M while the evaluated masks increase only from 189 to 305. Resource-weight changes alter depth/routing priorities with at most a 0.44-point mean validation shift.
\end{abstract}


\section{Introduction}
Parameterized quantum circuits (PQCs) provide a common substrate for
near-term quantum machine learning and variational algorithms
\cite{mitarai2018,benedetti2019,sim2019,cerezo2021,abbas2021}.
Their deployability, however, depends on more than predictive accuracy:
sparse coupling graphs require logical-to-physical mapping and can introduce
additional routing operations, two-qubit gates, and circuit depth
\cite{li2019,murali2019}. Circuit structure is therefore increasingly
optimized together with hardware constraints through architecture search,
mapping-aware optimization, and quantum/classical co-design
\cite{jiang2021,du2022,wang2022,huang2024,linghu2024}. In this setting, quantum neural networks (QNNs) are typically implemented as parameterized quantum
circuits whose trainable gate parameters are optimized for a learning task
\cite{mitarai2018,benedetti2019,abbas2021}.

This paper asks a different systems question: \emph{given an already trained QNN, which existing trainable entanglers should be removed before deployment?} Prior QNN compression and pruning work have shown that compilation awareness matters \cite{hu2022}, that iterative angle-based pruning can improve noise resilience \cite{sajadimanesh2025}, and that routing-aware removal of small controlled rotations can improve circuit fidelity \cite{escofet2026}. However, a local gate decision does not capture the non-additive effect of pruning a \emph{set} of interactions: removing several logical edges can change placement, SWAP insertion, and depth in ways not predicted by summing singleton savings.

\method treats post-training pruning as an accuracy-constrained, compiler-in-the-loop \emph{mask search}. It first combines task sensitivity and physical cost to reduce the candidate space; it then evaluates complete masks after transpilation and uses fixed-mask recovery only after structural selection. The held-out test split never participates in mask, optimizer, recovery, or resource-policy selection. \textit{\textbf{Our novel contributions are}}: 
\begin{itemize}
    \item \textbf{Compiler-aware post-training mask search.} We combine task saliency and leave-one-gate-out physical cost for screening, then evaluate complete fixed-cardinality masks after multi-topology transpilation rather than summing local gate scores.
    \item \textbf{Matched eight-seed evidence.} At 55\% pruning, search improves mean $\RT$ over the strongest simple baseline by 3.5/1.7/8.9~pp on Breast Cancer/Digits01/Iris, while direct recovery raises feasibility from 16/24 to 22/24 runs.
    \item \textbf{Deployment and scaling evidence.} Re-weighting the same evaluated masks changes depth/routing priorities with at most 0.44~pp mean validation shift, while a controlled 4--12q fixed-$k=4$ probe expands the raw space by $24{,}400\times$ but evaluated masks by only $1.62\times$.
\end{itemize}

\begin{figure*}
    \centering
    \includegraphics[width=\linewidth]{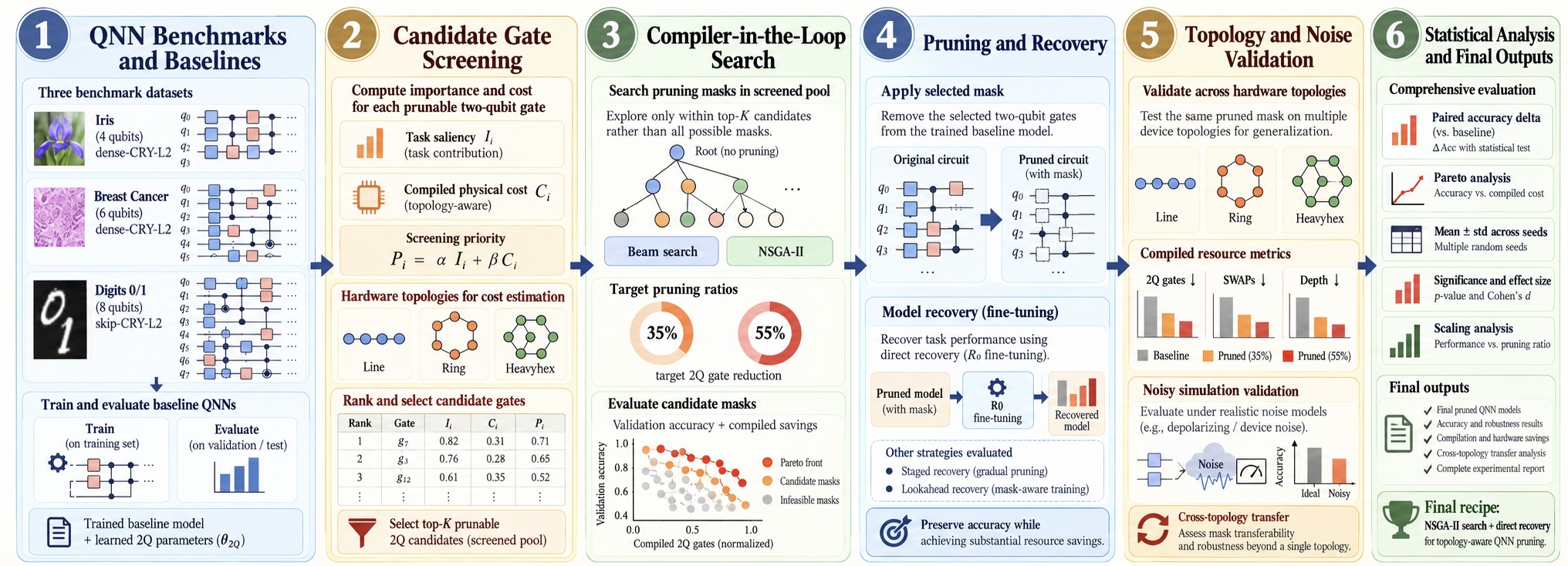}
    \caption{Overview of the ConQUESTA evaluation pipeline and main analyses}
    \label{fig:methodology}
\end{figure*}

\section{Related Work and Positioning}
QAS methods optimize circuit structure during model design, ranging from
task-aware architecture search \cite{du2022} and architecture/mapping
co-search \cite{wang2022,huang2024} to quantum/classical co-design
\cite{jiang2021}; QAS has also been demonstrated on a superconducting
processor \cite{linghu2024}. \method instead freezes the trained QNN as
the starting point and searches only masks over its existing trainable
entanglers.

Compiler-aware QNN compression is closer to our setting. CompVQC uses
compilation-aware pruning/quantization levels and ADMM-based compression
\cite{hu2022}; NR-QNN iteratively prunes small-angle gates and fine-tunes
the remaining circuit \cite{sajadimanesh2025}; and recent work removes
controlled rotations when their routing cost outweighs their utility for
fidelity \cite{escofet2026}. \method differs by combining task saliency and
compiler cost for screening, then searching \emph{joint fixed-cardinality
masks} using full-mask transpilation across multiple coupling maps.

\begin{table}[t]
\caption{Conceptual positioning relative to nearby optimization approaches.}
\label{tab:position}
\centering
\scriptsize
\setlength{\tabcolsep}{2.7pt}
\begin{tabular}{p{1.45cm}p{0.9cm}p{4.0cm}}
\toprule
Method & Start & Main distinction \\
\midrule
CompVQC \cite{hu2022} & QNN & compiler-aware compression levels + ADMM \\
NR-QNN \cite{sajadimanesh2025} & QNN & angle-threshold pruning + fine-tuning \\
QCP'26 \cite{escofet2026} & circuit & routing-aware local controlled-rotation pruning \\
QuantumNAS \cite{wang2022} & design & architecture + qubit-mapping co-search \\
\rowcolor{lightblue}
\textbf{\method} & trained QNN & task/compiler screening + joint mask search \\
\bottomrule
\end{tabular}
\end{table}

\section{Method}

\subsection{Task/Compiler Screening}
Consider $m$ trainable two-qubit entanglers $g_i$ with parameters $\theta_i$. Task importance is the first-order saliency
\begin{equation}
I_i=\mathbb{E}_{(x,y)}\!\left[\left|\theta_i\frac{\partial \mathcal{L}(x,y)}{\partial\theta_i}\right|\right].
\label{eq:saliency}
\end{equation}
For gate $i$, leave-one-gate-out transpilation yields normalized changes in native two-qubit count, routed SWAPs, and depth. For topology $b$ and transpiler seed $s$,
\begin{equation}
C_i^{(b,s)}=.60\frac{\Delta 2Q}{2Q_0}+.25\frac{\Delta S}{\max(S_0,1)}+.15\frac{\Delta D}{D_0},
\label{eq:gatecost}
\end{equation}
with negative compiler deltas clipped to zero. The robust gate cost takes the median across transpiler seeds followed by the 75th percentile across the scoring topologies. After min--max normalization,
\begin{equation}
P_i=\widehat C_i(1-\widehat I_i).
\label{eq:priority}
\end{equation}
Large $P_i$ means physically expensive but task-insensitive. Candidate-screening forms a diversified candidate pool from the joint priority, high physical cost, low task saliency, and a gradient-only joint view; the production pool is capped at $K=20$.

\subsection{Full-Mask Objective and Selection}
For a pruning mask $M$, compilation is performed on line, ring, and heavy-hex maps. Let $\widetilde r_{2q}^{b}$, $\widetilde r_{s}^{b}$, and $\widetilde r_{d}^{b}$ be the median fractional reductions over transpiler seeds for topology $b$. The default deployment policy is
\begin{equation}
\RT(M)=\frac{1}{|\mathcal B|}\sum_{b\in\mathcal B}\left(.60\widetilde r_{2q}^{b}+.25\widetilde r_s^{b}+.15\widetilde r_d^{b}\right).
\label{eq:t1}
\end{equation}
A mask is feasible when
\begin{equation}
\Aval(M)\geq\Aval(\varnothing)-\epsilon,\qquad \epsilon=0.01.
\label{eq:constraint}
\end{equation}
NSGA-II \cite{deb2002} performs fixed-cardinality search over validation accuracy, 2Q reduction, depth reduction, and SWAP reduction. Final selection maximizes $\RT$ among feasible masks; if none is feasible, it first minimizes accuracy violation and then maximizes $\RT$. The final recipe uses NSGA-II followed by direct fixed-mask fine-tuning. Validation data are used for mask and recovery selection; the held-out test split is evaluated only after the pruned structure is frozen. An eight-seed Beam comparison does \emph{not} establish optimizer superiority: NSGA-II vs. Beam gives $92.87\pm3.40$ vs. $92.54\pm3.35$ validation accuracy and $63.54\pm2.88$ vs. $61.65\pm3.43$ search-time resource score, with paired supporting $p=.56$ and $p=.22$, respectively. We therefore treat NSGA-II as the chosen implementation, not a separate algorithmic claim.

\section{Experimental Design}
The benchmarks are Iris (4-qubit dense-CRY, 12 prunable entanglers), Breast Cancer (6-qubit dense-CRY, 30 entanglers), and Digits 0-vs-1 (8-qubit skip-CRY). The statistical campaign uses model seeds $\{11,22,33,44,55,66,77,88\}$ at 35\% and 55\% pruning. All matched baselines start from the same trained checkpoint, prune to the same target cardinality, and receive the same direct-recovery budget. Specifically, direct fixed-mask recovery uses \textit{eight} fixed fine-tuning epochs at a learning rate of 0.008 (with early stopping disabled), compared with dense training for up to 35 epochs with patience 8. Because the structural mask remains fixed during recovery, this one-time offline fine-tuning only updates surviving parameters; it does not reintroduce removed entanglers or undo the compiled 2Q/SWAP/depth reductions. The one-shot \method comparator ranks eligible entanglers by Eq.~\eqref{eq:priority} and removes the top-$k$ gates directly, so its only structural difference from search-based \method is local ranking versus joint-mask optimization. Resource evaluation uses optimization level 3 and transpiler seeds 101/202/303. Search uses line/ring/heavy-hex; an older three-seed confirmation additionally evaluates an unseen grid topology. Noise stress testing uses 2048 shots on line/heavy-hex at four depolarizing settings from $(p_1,p_2)=(2\!\times\!10^{-4},2\!\times\!10^{-3})$ to $(2\!\times\!10^{-3},2\!\times\!10^{-2})$.

We report \textit{mean $\pm$ sample standard deviation} across model seeds. For paired method comparisons we additionally report the median paired difference, bootstrap median confidence interval, Wilcoxon signed-rank $p$ as supporting rather than confirmatory evidence, and rank-biserial effect size; $n=8$ remains a small-sample regime.

For the scaling-only experiment, we use a controlled dense-CRY L2 family at $q\in\{4,6,8,10,12\}$ with fixed $k=4$, thereby isolating compiler/search cost from application scaling. It reports raw candidates $N$, screened pool $K$, $\binom Nk$, unique evaluated masks $B$, transpilation counts, wall-clock, and peak RSS.

\begin{table}[t]
\caption{Research questions guiding the evaluation.}
\label{tab:rqs}
\centering
\footnotesize
\setlength{\tabcolsep}{3pt}
\renewcommand{\arraystretch}{1.08}
\begin{tabularx}{\columnwidth}{>{\bfseries}p{0.55cm}X}
\toprule
RQ & Research Question \\
\midrule
RQ1 & Does compiler-aware joint-mask search improve over simpler
post-training pruning at matched sparsity and recovery budget? \\

RQ2 & How effectively does fixed-mask recovery restore the
validation-accuracy constraint after aggressive pruning? \\

RQ3 & How sensitive are the selected masks and resource savings
to the deployment resource policy? \\

RQ4 & Does candidate screening limit combinatorial search growth
as the number of prunable entanglers increases? \\
\bottomrule
\end{tabularx}
\end{table}

\section{Results}

\subsection{RQ1: Joint-Mask Search vs. Simple Pruning}

\begin{figure}[t]
\centering
\includegraphics[width=\columnwidth]{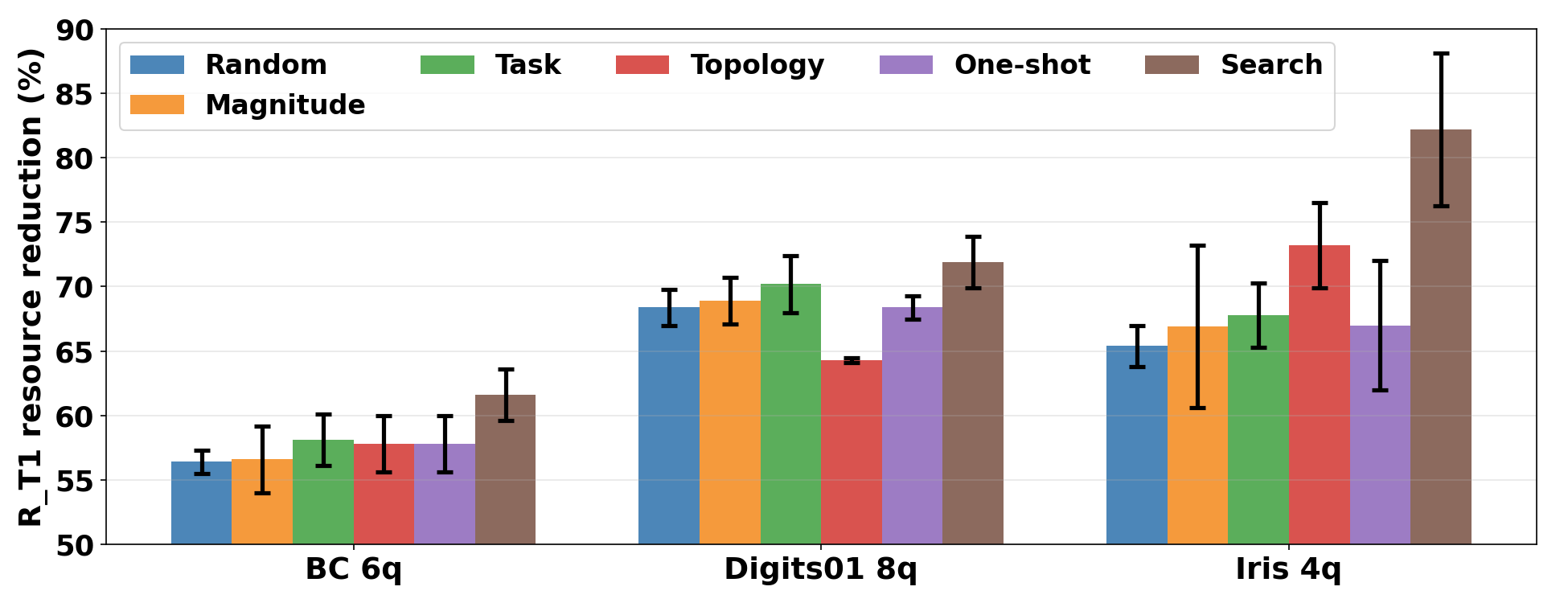}
\caption{Matched 55\% pruning: mean $\RT$ resource reduction $\pm$ sample SD over eight model seeds. ``One-shot'' applies Eq.~\eqref{eq:priority} directly without combinatorial mask search.}
\label{fig:baseline}
\end{figure}

Figure~\ref{fig:baseline} and Table~\ref{tab:main} compare methods at
\textbf{55\% pruning} under the same checkpoint, sparsity, and recovery
budget. Joint-mask search achieves the highest mean $\RT$ on all three
benchmarks. The strongest simple baseline is task-only on Breast Cancer and Digits01, and compiler-only on Iris; search improves mean $\RT$ by $+3.46\pp$, $+1.74\pp$, and $+8.89\pp$, respectively. Accuracy is deliberately not presented as uniformly superior: on Digits01, magnitude and task-only pruning match search validation accuracy, while on Iris compiler-only is $0.42\pp$ higher. The supported claim is therefore \emph{better compiled physical efficiency at comparable task accuracy}, not universal accuracy dominance.

Against one-shot \method specifically, the paired median resource-score gain of search is $4.88\pp$ on Breast Cancer (95\% bootstrap median CI $0.01$--$7.07\pp$, supporting $p=.023$, $|r_{rb}|=.89$), $2.74\pp$ on Digits01 (CI $1.85$--$5.45\pp$, $p=.008$, $|r_{rb}|=1.00$), and $16.78\pp$ on Iris (CI $15.52$--$18.89\pp$, $p=.016$, $|r_{rb}|=.94$). This isolates the value of mask-level search beyond the same screening heuristic.

\begin{table}[t]
\caption{Eight-seed results at \textbf{55\% pruning}. $\RT$ and validation columns are mean $\pm$ sample SD (\%). ``Best simple'' is selected by mean $\RT$ from random/magnitude/task/compiler-only.}
\label{tab:main}
\centering
\scriptsize
\setlength{\tabcolsep}{2.0pt}
\resizebox{\columnwidth}{!}{%
\begin{tabular}{llccccc}
\toprule
Benchmark & Best simple & Simple $\RT$ & Search $\RT$ & Gain & Recovered val. & Feasible \\
\midrule
BC 6q & Task-only & $58.10\pm2.00$ & $\mathbf{61.55\pm2.10}$ & $\mathbf{+3.46}$ & $94.08\pm3.17$ & $5/8\to8/8$ \\
Digits01 8q & Task-only & $70.23\pm2.20$ & $\mathbf{71.97\pm1.99}$ & $\mathbf{+1.74}$ & $99.31\pm0.74$ & $5/8\to6/8$ \\
Iris 4q & Compiler-only & $73.31\pm3.22$ & $\mathbf{82.21\pm5.87}$ & $\mathbf{+8.89}$ & $87.08\pm6.28$ & $6/8\to8/8$ \\
\bottomrule
\end{tabular}%
}
\end{table}

\subsection{RQ2: Accuracy-Constraint Recovery}
At aggressive \textbf{55\% pruning}, direct recovery increases feasibility from $16/24$ to $22/24$ seed--benchmark runs (66.7\% to 91.7\%). Across both 35\% and 55\%, feasibility changes from $39/48$ to $45/48$. Recovery does not imply universal held-out gains: at 55\%, mean test-accuracy deltas relative to the dense model are $+0.88\pp$ (BC), $-0.52\pp$ (Digits01), and $-1.25\pp$ (Iris), each with substantial seed variance. The appropriate claim is therefore accuracy \emph{constraint recovery}, not test-set improvement. The two remaining Digits01 failures are seeds 55 and 77: both recover to
98.61\% validation accuracy against a 99.0\% feasibility threshold, missing
the constraint by only 0.39~pp.

\subsection{RQ3: Deployment-Policy Sensitivity}

The default weights in Eq.~\eqref{eq:t1} are a deployment policy, not a universal optimum. We re-score the \emph{same stored masks} using uniform $(1/3,1/3,1/3)$, depth-oriented $(.40,.20,.40)$, and routing-oriented $(.40,.40,.20)$ policies. The offline implementation exactly reproduces the original default selection in all 48 benchmark--seed--ratio conditions before changing any weight. The policy knob behaves as intended: depth weighting buys additional depth reduction and routing weighting increases SWAP reduction, while the largest mean validation shift across all alternatives is only $0.44\pp$. At 55\%, masks are especially stable ($J\ge .93$ for BC, $.99$ for Digits01, and $.94$ for Iris). Depending on condition/policy, 87.5--100\% of selected masks lie on the Pareto frontier computed from aggregate accuracy/2Q/depth/SWAP summaries; the remaining cases reflect that the actual T1 selector scalarizes topology-wise median reductions rather than the global aggregate columns.

\begin{table}[t]
\caption{Weight sensitivity. Entries are 35\%/55\%; $J_{\min}$ is minimum mean Jaccard similarity to the default mask, and $\Delta$ values are changes from default in percentage points.}
\label{tab:weights}
\centering
\scriptsize
\setlength{\tabcolsep}{3.1pt}
\begin{tabular}{lcccc}
\toprule
Benchmark & $J_{\min}$ & Depth-pol. $\Delta D$ & Route-pol. $\Delta S$ & max $|\Delta A|$ \\
\midrule
BC & $.70/.93$ & $+1.63/+1.53$ & $+1.74/+0.79$ & $0.44/0.11$ \\
Digits01 & $.72/.99$ & $+3.82/+0.12$ & $+0.33/0.00$ & $0.00/0.00$ \\
Iris & $.95/.94$ & $+0.00/+1.37$ & $+0.69/0.00$ & $0.00/0.00$ \\
\bottomrule
\end{tabular}
\end{table}

\subsection{RQ4: Does Screening Avoid Combinatorial Growth?}
\begin{figure}[t]
\centering
\includegraphics[width=\columnwidth]{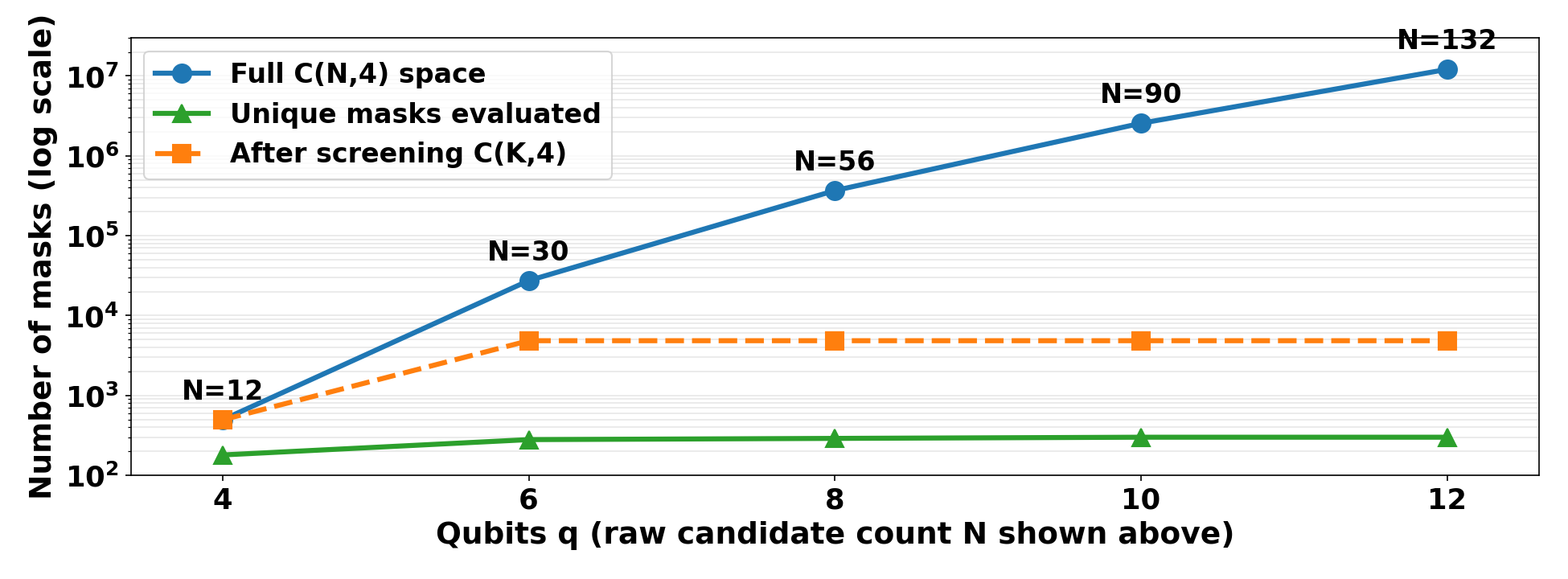}
\caption{Scaling study at fixed $k=4$. Screening bounds the combinatorial domain while the raw full-mask space grows by four orders of magnitude.}
\label{fig:scaling}
\end{figure}

Figure~\ref{fig:scaling} isolates the combinatorial effect in a \textbf{controlled synthetic dense-CRY fixed-$k=4$ probe}; it is not a fixed-ratio application benchmark. Raw prunable entanglers grow from $N=12$ to $132$, so the full mask space grows from $495$ to $12{,}082{,}785$ ($24{,}400\times$ increase). Screening caps $K$ at 20 from 6 qubits onward, while unique evaluated masks grow only from $189$ to $305$ ($1.62\times$ increase). Consequently, the explored fraction falls from 38.18\% to 0.0025\%. Search time grows from 265 to 3254~s and time per unique mask from 1.40 to 10.66~s, showing that screening suppresses combinatorial growth while per-mask compilation becomes the next bottleneck. Compile-cache hit rate is zero, so this trend is attributable to screening and bounded search rather than cache reuse. At a fixed 55\% ratio, $k$ would grow with $N$ (e.g., beyond the current $K=20$ pool at $N=56$ and $132$), so fixed-ratio scaling remains outside the claim of this experiment.

\begin{table}[t]
\caption{Scaling cost. $B$ is the mean unique masks evaluated; RSS is mean peak resident memory.}
\label{tab:scaling}
\centering
\scriptsize
\setlength{\tabcolsep}{2.7pt}
\begin{tabular}{rrrrrrr}
\toprule
$q$ & $N$ & $K$ & $\binom{N}{4}$ & $B$ & $B/\binom{N}{4}$ & s/mask \\
\midrule
4  & 12  & 12 & 495       & 189 & 38.18\% & 1.40 \\
6  & 30  & 20 & 27,405    & 290 & 1.058\% & 2.87 \\
8  & 56  & 20 & 367,290   & 298 & 0.0812\% & 4.72 \\
10 & 90  & 20 & 2,555,190 & 303 & 0.0119\% & 7.25 \\
12 & 132 & 20 & 12,082,785& 305 & 0.0025\% & 10.66 \\
\bottomrule
\end{tabular}
\end{table}

Screening transpilation calls grow from 39 to 399 and search calls from
roughly 570 to 919, while peak RSS rises from 0.79 to 1.55~GB.

The earlier three-seed grid and noise experiments are secondary stress
tests because they were not repeated in the eight-seed campaign.
Masks selected over line/ring/heavy-hex retain substantial savings on the
unseen grid (Table~\ref{tab:secondary}), suggesting transfer beyond the
exact search maps but not invariance to arbitrary device layouts.
Noise effects are benchmark-dependent; we therefore treat \method as a
resource-efficiency method rather than an error-mitigation technique.

\begin{table}[t]
\caption{Secondary three-seed 55\% stress tests. Grid columns are mean reductions (\%); noise is the range of recovered-minus-dense accuracy over four tested noise levels and two search topologies.}
\label{tab:secondary}
\centering
\scriptsize
\setlength{\tabcolsep}{3.1pt}
\begin{tabular}{lrrrr}
\toprule
Benchmark & Grid 2Q & Grid depth & Grid SWAP & Noise $\Delta A$ (pp) \\
\midrule
BC & 54.4 & 46.8 & 72.7 & $-0.8$ to $+0.8$ \\
Digits01 & 65.6 & 71.1 & 86.7 & $+0.9$ to $+2.8$ \\
Iris & 79.5 & 73.8 & 100.0 & $-3.3$ to $-2.2$ \\
\bottomrule
\end{tabular}
\end{table}

\section{Discussion \& Conclusion}

\method provides three main systems-level findings. \textit{\textbf{First}}, a joint compiler-aware
mask search improves $\RT$ over matched simple pruning baselines at 55\% sparsity by 3.46/1.74/8.89~pp on BC/Digits01/Iris while maintaining comparable accuracy. \textbf{\textit{Second}}, the resource scalarization acts as a deployment policy rather
than a tuned optimum: alternative weights shift depth and routing savings while
changing mean validation accuracy by at most 0.44~pp. \textbf{\textit{Third}}, candidate-screening substantially limits combinatorial growth in the controlled fixed-$k$ scaling
study: from 4 to 12 qubits, the raw mask space grows $24{,}400$ times, whereas
the number of evaluated masks grows only $1.62$ times.

This efficiency is obtained at a one-time offline cost. Estimated NSGA-II search
times average approximately 4.2, 46.1, and 22.2 min for Iris, BC, and Digits01,
respectively, compared with dense-training times of about 0.12, 13.2, and
8.7 min. Direct fixed-mask recovery uses only eight additional fine-tuning
epochs and does not alter the selected structure or undo the compiled
2Q/SWAP/depth reductions.

%The conclusions remain bounded by the 4--8-qubit production models, the
%fixed-$k$ nature of the 10--12-qubit scaling probe, three-seed topology/noise
%stress tests, lack of real-QPU execution, and the absence of an independent
%search-stochasticity sweep. We therefore do not claim fixed-ratio large-QNN
%scalability, arbitrary-backend invariance, or hardware-level fidelity.
Overall, \method occupies a distinct point between local post-training pruning
and full QAS: it restricts search using task/compiler information, evaluates
complete masks after transpilation, and freezes the selected structure before
recovery and held-out test evaluation. Evidently, our results position compiler-aware
mask search as a practical post-training complement to quantum
architecture search rather than a replacement for it.

\section{Acknowledgement}

This work was supported by the NYUAD Center for Quantum and Topological Systems (CQTS), funded by
Tamkeen under the NYUAD Research Institute grant CG008. This research was carried out on the High Performance Computing resources at New York University Abu Dhabi.

\balance
\clearpage
\bibliographystyle{IEEEtran}
\bibliography{refs}

\end{document}